\documentclass{raa}  
\usepackage{graphicx,times}
\usepackage{txfonts}

\usepackage{natbib}

\bibpunct{(}{)}{;}{a}{}{,}

\usepackage[pagebackref=true]{hyperref}

\usepackage{lipsum}
\usepackage{subcaption}         
\usepackage{lscape}             
\usepackage{placeins}           

\begin{document}

\title{A New Cartwheel-like Collisional Ring Galaxy
}

\volnopage{ 2026 Vol.\ 26 }
\setcounter{page}{1}

\author{Vladimir P. Reshetnikov
       \inst{1,2}
\and Sergey S. Savchenko
       \inst{1,2} 
\and Alexander A. Marchuk
       \inst{1,2}
\and Ilia V. Chugunov
       \inst{1,3} 
\and Matvey D. Kozlov
       \inst{1,2} 
\and Dmitry I. Makarov
       \inst{1,4}
\and Aleksandra V. Antipova
       \inst{1,4} 
\and Anastasia M. Sypkova
       \inst{1,2}
          }

\institute{Pulkovo Astronomical Observatory, Russian Academy of Sciences, St.Petersburg 196140, Russia\\
\and 
Saint Petersburg State University, 7/9 Universitetskaya nab., St.Petersburg 199034, Russia; {\it v.reshetnikov@spbu.ru}\\
\and
Sternberg Astronomical Institute, Lomonosov Moscow State University, Universitetsky Pr. 13, 119234 Moscow, Russia\\
\and
Special Astrophysical Observatory, Russian Academy of Sciences, Nizhnii Arkhyz 369167, Russia\\
\vs\no
{\small Received~~20xx month day; accepted~~20xx~~month day}}


\abstract
{We report the discovery of a new Cartwheel-type collisional ring galaxy, PGC\,1112751, which we named
``Eridanus Wheel'' (EW).  Such systems result from head-on collisions between galaxies and are of considerable
interest as laboratories for studying star formation in propagating density waves and the response of the star-gas
disk of galaxies to strong external perturbations. During a systematic visual inspection of fields from the DESI
Legacy Imaging Surveys, we identified a galaxy at a redshift of $z=0.0856$ whose morphology closely resembles that of
the famous Cartwheel Galaxy. EW exhibits a well-defined inner ring and a more diffuse outer ring, with so-called
``spokes'' visible in the region between them. The ring galaxy and its possible intruder are connected by a faint
optical bridge. The projected distance between the centers of the galaxies is about 60 kpc.  Using data from the DESI
Legacy Surveys, we performed a photometric study of galaxies in the $griz$ filters. We conclude that the Eridanus Wheel
is a giant late-type galaxy with a strong radial color gradient, whose observed morphology is most likely explained by
a relatively recent head-on collision with an early-type galaxy. The further evolution of this object will most likely
lead to the formation of a galaxy with a low surface brightness disc.  
}
               
\titlerunning{Cartwheel-like galaxy}
\authorrunning{V. P. Reshetnikov, S. S. Savchenko, A. A. Marchuk et al.}
\maketitle


%

\section{Introduction}

One of the most spectacular manifestations of interactions between galaxies is the formation of so-called collisional
ring galaxies (CPGs; see numerous examples in \citealt{madore2009}).  Such systems form during nearly head-on collisions
between two galaxies, in which one (the ``intruder'') travels at high velocity through the disk of a spiral galaxy (the
``target'') approximately along its rotation axis. This encounter generates propagating density waves in the target
galaxy, leading to the formation of a distinguishing ring-like morphology (e.g. \citealt{lt1976}; \citealt{ts1977}).
CRGs are very rare objects (e.g., \citealt{few1986}; \citealt{sr2022}) and only a handful of
them have been studied in detail (e.g., \citealt{charm1993}; \citealt{wallin1994}; \citealt{romano2008};
\citealt{conn2011}; \citealt{mondal2025}).

Perhaps the best-known example is the Cartwheel Galaxy (\citealt{zwicky1941}), usually regarded as the prototype of this
phenomenon (\citealt{as1996}). This system has been studied extensively, both observationally
(e.g., \citealt{higdon1996}; \citealt{amram1998}; \citealt{barway2020}; \citealt{ditrani2024}) and through numerical
simulations (\citealt{sm1993}; \citealt{hern1993}; \citealt{horellou2001}; \citealt{renaud2018}).

The study of CRGs is interesting for many reasons. For example they serve as laboratories for the study of density wave
triggered star formation and consequent stellar evolution, expanding rings provide responsive probes of both the shape
and strength of the gravitational potential of the targeted galaxy (e.g., \citealt{as1996}). \cite{mapelli2008} suggested
that CRGs could be the progenitors of giant low surface brightness galaxies. According to their numerical simulations,
at a late evolutionary stage ($\geq 0.5$ Gyr), the expanding ring fades, its size reaches $\sim10^2$ kpc, and it
begins to resemble the disks of low surface brightness galaxies.

In this note, we describe a newly identified collisional ring galaxy, with a morphology similar to that of the Cartwheel
Galaxy. This is the galaxy PGC\,1112751, located in the constellation Eridanus. Following the example of
\cite{conn2011} article, we designate it the Eridanus Wheel.


\section{Discovering of Eridanus Wheel}

The unusual morphology of PGC\,1112751 galaxy was discovered by chance during the preparation of a new sample of edge-on
galaxies from the DESI Legacy Imaging Surveys\footnote{https://www.legacysurvey.org/} (\citealt{desi}). This sample of
edge-on galaxies (Marchuk, 2026, in preparation) was constructed automatically using an artificial neural network
classifier. To exclude false positive classifications, the entire sample ($\sim 190000$ images covering a total area of
107 square degrees) was inspected visually. During this inspection, PGC\,1112751 was spotted near an edge-on candidate
UGC\,02650.

Figure\,1 shows the field around the edge-on spiral galaxy UGC\,02650 at a redshift of $z=0.028$ (UGC\,02650 is located to the
right and slightly above the center of the field). The object of interest -- the Eridanus Wheel (EW) -- is located
approximately 1\farcm7 to the northeast at $z=0.0856$. EW and its companion (a small galaxy half a minute south --
Figure\,2) are visible in the background of a group of galaxies at a redshift of $z=0.028$ (Fig\,1).

\begin{figure}
\centering
\includegraphics[width=8.7cm, angle=0, clip=]{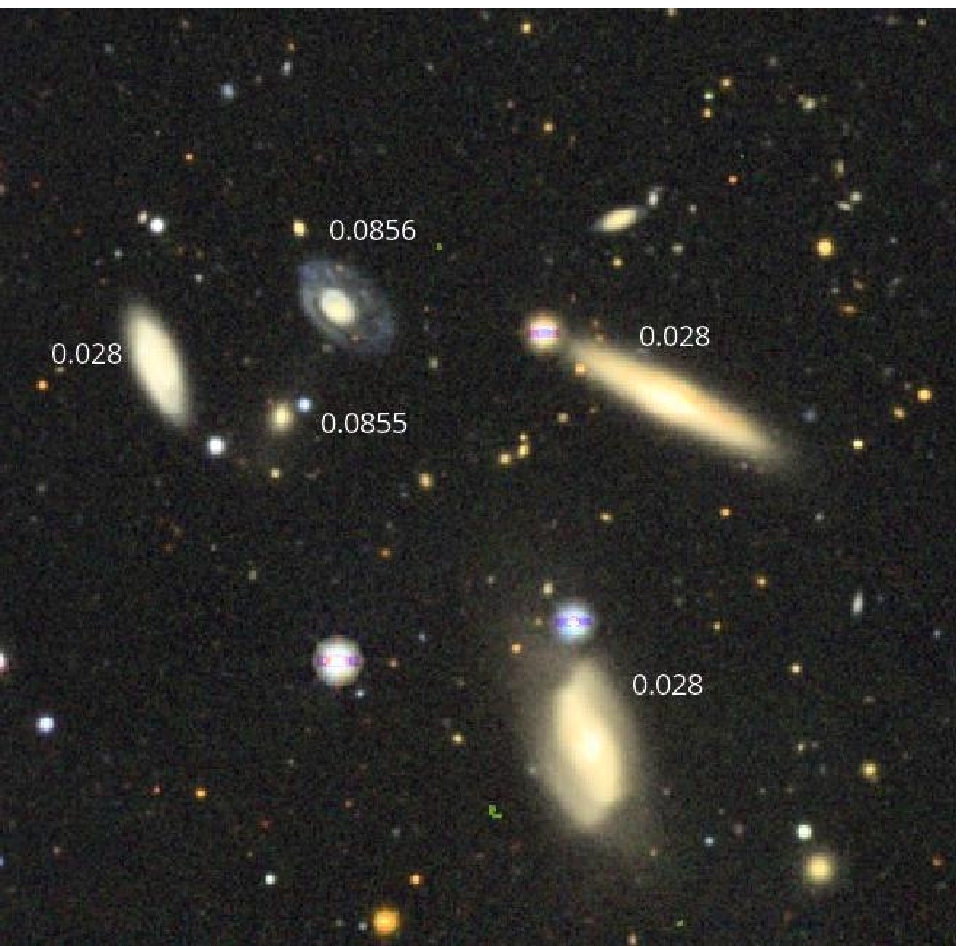}
\caption{DESI Legacy Surveys DR10 image of the field around UGC\,02650 (see the text). The numbers near the galaxies 
indicate their redshifts according to the NASA/IPAC Extragalactic Database (NED) and DESI DR1 (\citealt{abdul2025}). 
North is up and east is to the left. The image size is 4\farcm7 $\times$ 4\farcm7.}
\end{figure}

The galaxy has a bright, clearly visible inner ring and a more diffuse and non-uniform outer ring.  The centers of the
rings are offset relative to each other.  Weak curved structures, similar to the ``spokes'' of the Cartwheel Galaxy, are
clearly visible in the space between the two rings (Figure\,2). The possible intruder, WISEA\,J031742.00-014147.3, is
connected to EW by a weak optical bridge. The intruder exhibits a disturbed outer morphology, consistent with a recent
interaction with the host galaxy (Figure\,2).

Within the standard flat $\Lambda$CDM cosmological model ($\Omega_m=0.3$, $\Omega_{\Lambda}=0.7$, $H_0=70$
km\,s$^{-1}$\,Mpc$^{-1}$), we adopt the luminosity distance to the galaxy $D_L = 390$\, Mpc and the scale
1$\farcs$60/kpc.

\section{Photometric characteristics}

\begin{figure}
\centering
\includegraphics[width=4.7cm, angle=0, clip=]{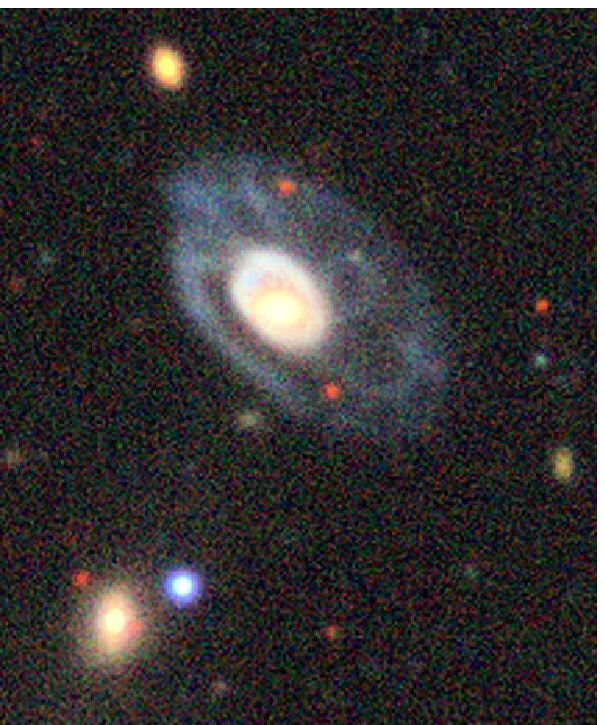} \\
\includegraphics[width=6.05cm, angle=-90, clip=]{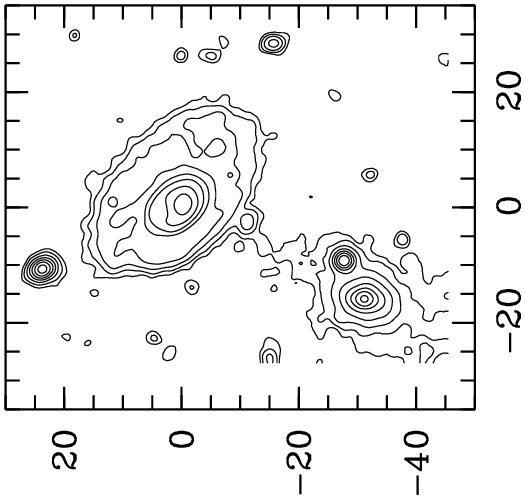}~~~~~~~~~~~
\caption{{\it Top}: Original DESI Legacy Surveys DR10 image of Eridanus Wheel.
The distance between the centers of EW and the companion (small galaxy in the lower left corner 
of the image) is 37$\arcsec$.
{\it Bottom}: Smoothed isophotes from the combined $griz$ image of the same field. Both axes are in
arcseconds, isophotes step is 0.$^m$75/$\Box''$.
}
\end{figure}

To analyze the photometric structure of the galaxies, we performed a photometric decomposition of the DESI Legacy
Imaging Surveys DR10 data in the $griz$ bands using the {\small IMFIT} package \citep{erwin2015}. The decomposition
model included four components: a S\'ersic function representing a central component (bulge), two Gaussian rings and an
additional S\'ersic component with low $n$-index to describe the underlay of a galaxy (mainly the light between the two
rings). The point spread function was constructed according to the DESI Legacy recommendations
\footnote{\url{https://www.legacysurvey.org/dr10/psf/}}. The images in the DESI Legacy Survey are already
sky-subtracted, but we checked a possible remnants of the background light by including the \textit{FlatSky} function
in our model. The average background level turned out to be $\approx 29^m/\Box''$ in the $r$-band, and we neglected it
in the further analysis. To estimate uncertainties of the decomposition parameters we used bootstrap option of the
{\small IMFIT} package with 1000 runs for each band.
The result of the modeling in the $g$ band is shown in Figure\,3.

\begin{figure}
\centering
\includegraphics[width=9cm, angle=0, clip=]{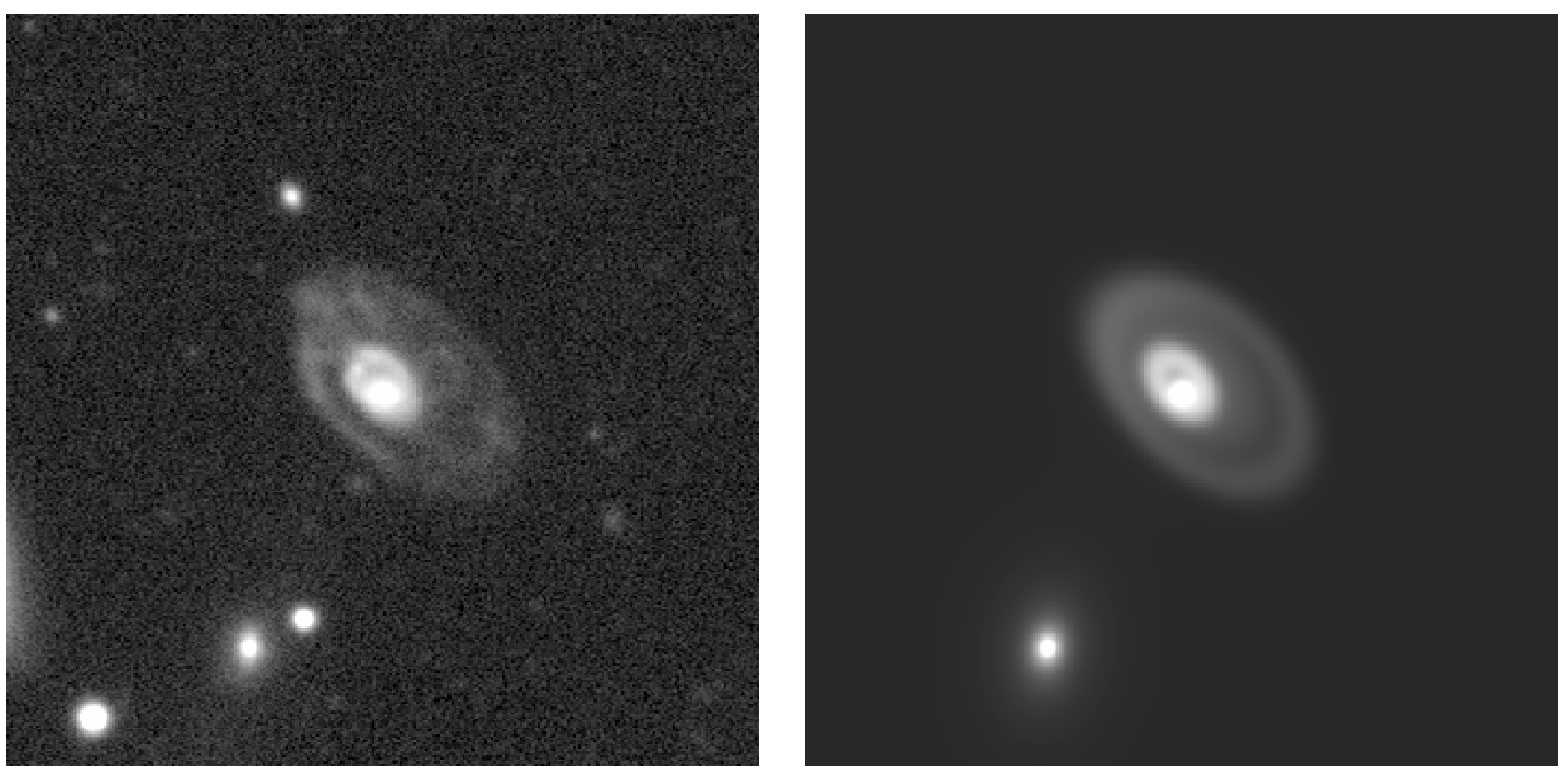}
\caption{Decomposition results in the $g$ filter. Left: original image, right: 2D model.}
\end{figure}

The integral characteristics of galaxies derived from the decomposition are summarized in Table\,1.  Diameters and
apparent axial ratios were determined by outer isophotes constructed from the combined $griz$ images (Figure\,2,
bottom). The peripheral regions of the probable intruder are distorted, and therefore its parameters are determined in
part arbitrarily.  As seen in the table, EW is a giant galaxy whose color indices are typical of late-type spiral
galaxies; the companion is approximately half as massive as EW, and its colors are more typical of early-type galaxies
(e.g., \citealt{fukugita}).

\begin{table}
\caption{General characteristics of the two galaxies}
\begin{center}
\begin{tabular}{|c|c|c|}
\hline
          & Eridanus Wheel                    & Companion \\
\hline
$\alpha$(2000)$^a$ & 03$^h$ 17$^m$ 40.$^s$93      &  03$^h$ 17$^m$ 42.$^s$01         \\
$\delta$(2000)$^a$ & --01$^{\circ}$ 41$'$ 16.$''$2 &  --01$^{\circ}$ 41$'$ 47.$''$4    \\                 
\hline
$z^b$          & 0.0856                       &  0.0855                       \\
$r_0^c$          &  16.26                       &  17.36                        \\
$(g-r)_0$      &  +0.57                       &  +0.82                        \\
$(r-i)_0$      &  +0.14                       &  +0.43                        \\
$(i-z)_0$      &  +0.34                       &  +0.26                        \\
$D$ (kpc)      &  60                          &  20:                          \\
$b/a$          &  0.6                         &  0.8:                         \\
$M_{r_0}$          &  $-21.7$                      &  $-20.6$                        \\
M$_*$ (M$_{\odot}$)$^d$  &  (6.6$\pm$1.6) $\times$ 10$^{10}$ & (3.2$\pm$0.8) $\times$ 10$^{10}$        \\   
\hline
\end{tabular}

\vspace{2mm}

\parbox[t]{150mm}
{\small
$^a$  NASA/IPAC Extragalactic Database (NED), 

$^b$  DESI DR1 (\citealt{abdul2025}),

$^c$  All the magnitudes, colors and luminosities in the paper are corrected 

for the Milky Way absorption,

$^d$  \cite{ebr2025}

}
\end{center}
\end{table}

\subsection{Eridanus Wheel}

\begin{figure}
\centering
\includegraphics[width=8cm, angle=0, clip=]{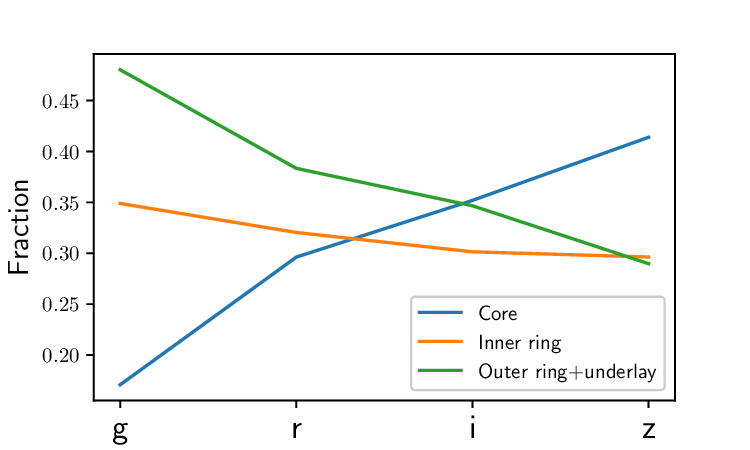}
\caption{The relative contribution of different subsystems to the total luminosity of the galaxy.
}
\end{figure}

Figure\,4 shows the contributions of different subsystems to the total EW luminosity as a function of filter. In the
shortest wavelength band ($g$), the main contribution to the galaxy's luminosity comes from the outer extended ring,
while the contribution of the central component (bulge) is small. With increasing wavelength, the relative contributions
of the two rings decrease, and in the $z$ band they become smaller than that of the bulge. Figure\,4 illustrates a strong
color gradient in EW:

\hspace{2cm} $(g-r)_0 = +1.17$ (bulge),

\hspace{2cm} $(g-r)_0 = +0.49$ (inner ring),

\hspace{2cm} $(g-r)_0 = +0.33$ (outer ring).\\
Such distribution of color indices by component is usual for CRGs (\citealt{as1996}).

\begin{figure*}
\centering
\includegraphics[width=10cm, angle=-90, clip=]{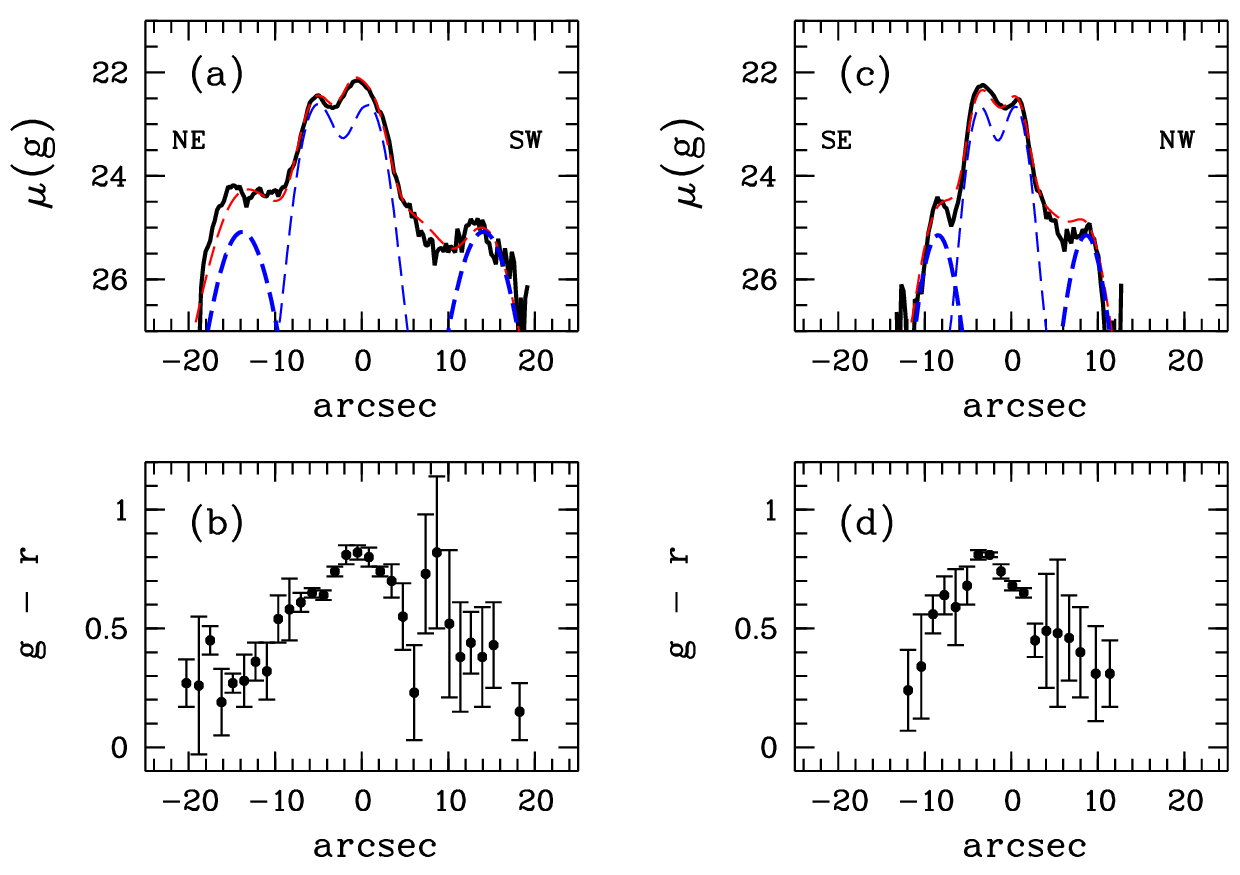}
\caption{Distributions of surface brightness in the $g$ band and color 
index $g-r$ along the major -- (a), (b) -- and minor -- (c), (d) -- axes of the outer ring.
The zero point for measuring distances corresponds to the center of the outer ring.
The black lines in (a) and (c) show the observed distributions, the red lines show the 
final result of photometric decomposition, and the blue dashed lines give the contributions of the 
inner (thin line) and outer (thick line) rings based on the modeling.
}
\end{figure*}

Figure\,5 displays the surface brightness profiles of EW along the major (P.A.=45$^\circ$) and minor
(P.A.=135$^\circ$) axes. The outer ring is very faint, with a surface brightness of $\mu(g) \approx (25-25.5)^m/\Box''$.
As it continues to evolve, the outer ring and the region of EW between the rings will apparently fade, and the 
galaxy as a whole may become similar to a giant low surface brightness galaxy.
We also note a strong color gradient along the major and minor axes of the galaxy, amounting to 
$\sim 0.02^m$/kpc (Figure\,5).

The characteristics of the galaxy bulge (core in Figure\,4) show a dependence on wavelength: in the $g$ filter, the
effective radius and S\'ersic index of the bulge are $r_e(g)=0\farcs9\pm 0\farcs1$ ($1.44\pm 0.2$ kpc) and
$n(g)=1.1\pm 0.1$, while in the $z$ filter, they increase to $1\farcs4\pm 0\farcs1$ ($2.24\pm 0.2$ kpc) and 2.1,
respectively. A possible reason for such a strong wavelength dependence is that there is a bar in the galaxy, whose
contribution increases in redder filters.  Analysis of the original images confirms this assumption. The central part of
the galaxy is slightly elongated, which can be explained by the presence of a bar with a diameter of 4$''$--5$''$ (6--8
kpc).  Thus, we tentatively conclude that EW hosts a pseudo-bulge and a bar. In this respect, EW resembles the Cartwheel
Galaxy (\citealt{barway2020}).

\subsection{Probable intruder}

Based on its morphology and color indices, the intruder looks like an early-type (E/S0) galaxy.  The galaxy is
surrounded by a very faint, extended envelope (Figure\,2). Using a single-component S\'ersic model, we found that its
effective radius is $4\farcs5 \pm 0\farcs6$ (average value across four filters) or about 7 kpc.  The value of the S\'ersic
index depends on the filter: in the $g$ band, it is equal to 6, while in other filters, the index is close to 10 or even
higher. Such high S\'ersic indices indicate a very extended, shell-like surface brightness distribution in the galaxy.

The galaxy is connected to EW by a faint bridge with surface brightness in the $g$ filter of $\mu (g) \sim 27^m/\Box''$
(Figure\,2). On the opposite (southern) side of the galaxy, a faint tidal feature is also visible.
Its brightness is comparable to, or slightly brighter than, that of the bridge connecting the galaxies (Figure\,2).

\subsection{Two rings}

Like the Cartwheel Galaxy, EW exhibits two ring structures. Both rings are relatively blue (sect.\,3.1).  The inner ring
is well defined, while the outer ring appears more heterogeneous and blurred, with variable thickness. Based on our
decomposition, we found that the radius of the inner ring does not depend on the filter and is equal to
$R_{inn} = 3\farcs4 \pm 0\farcs1$ ($5.4 \pm 0.2$ kpc). The width of the ring (defined as the standard deviation of the
Gaussian approximation) is $1\farcs3 \pm 0\farcs1$ or about 2 kpc.

The radius of the outer ring, as it turned out, depends on the filter: it varies from $14\arcsec \pm 0\farcs1$ in the
$g$ filter to $13\farcs1\pm 0\farcs 1$ in the $z$ filter.  The average radius for the four filters is
$R_{out} = 13\farcs5 \pm 0\farcs4$ (21.6 kpc).  Thus, the outer ring of EW is among the largest known structures of this
type. For example, according to \cite{struck2010}, the average diameter of the outer collisional rings is 28 kpc (see
Tables 1 and 2 in his work). The width of the outer ring is greater than that of the inner ring and is approximately
$2''\pm 0\farcs 1$ or about $3.2\pm 0.2$ kpc.

Another interesting parameter is the ratio of the radii of the two rings.  According to the analytical model of the
formation of collisional rings in galaxies (see the discussion and references in \citealt{struck2010}), the ratio of
sizes beween the first- and second-created rings should be 3.0. At first glance, this model appears to be consistent
with observations. Thus, according to data from \cite{struck2010}, for 15 galaxies with measured values of $R_{inn}$ and
$R_{out}$, the average value $\langle R_{out}/R_{inn} \rangle = 3.0 \pm 0.7$.  For EW, this value is 4.0, and for
the Cartwheel Galaxy, it is 4.2 (\citealt{struck2010}).  This means that some initial assumptions of the analytical model (impulse
approximation, collision symmetry, flat rotation curve, etc.) may not be satisfied in the cases of EW and the Cartwheel
Galaxy.

\section{Discussion and conclusions}

In this note, we present a newly identified collisional ring galaxy morphologically similar to the Cartwheel Galaxy.
The Eridanus Wheel galaxy was discovered incidentally during the visual inspection of fields selected from the DESI
Legacy Imaging Surveys.  The total surveyed area amounted to 107 $\Box^\circ$. This implies that, based on a rough
scaling, several hundred similar objects may exist over the entire celestial sphere in the nearby Universe.  Modern
catalogs contain $\approx10^2$ good candidates for CRGs in the nearby Universe (for example, \citealt{madore2009}),
suggesting that many additional systems remain to be identified.

It is likely that the galaxy has a pseudo-bulge and a bar inherited from the pre-collisional progenitor
of the EW. In this respect, Eridanus Wheel resembles the Cartwheel Galaxy (\citealt{barway2020}). 
The galaxy has two rings: an inner ring with a diameter of about 11 kpc and an outer ring with a diameter of about 43 kpc.
The outer ring and the inter-ring region have low surface brightness (Figure\,5), so that in the course of subsequent 
evolution they may fade and become indistinguishable from low surface brightness galactic disks.

The probable intruder has half the stellar mass of EW (Table\,1) and, based on its photometric characteristics,
resembles an early-type galaxy. The central regions of the galaxies are connected by a faint optical bridge.  In
addition, the intruder exhibits a counter-directed tidal tail and an extended envelope surrounding it (Figure\,2).  This
may indicate a relatively slow collision between galaxies and even that the galaxies may subsequently merge.  To
determine the future fate of this interesting interacting system, new detailed observational data and detailed dynamic
modeling are needed.

\begin{acknowledgements}
This work was supported by the Russian Science Foundation (project no. 24-72-10084). 

The DESI Legacy Imaging Surveys consist of three individual and complementary projects: the Dark Energy Camera Legacy 
Survey (DECaLS), the Beijing-Arizona Sky Survey (BASS), and the Mayall z-band Legacy Survey (MzLS). DECaLS, BASS 
and MzLS together include data obtained, respectively, at the Blanco telescope, Cerro Tololo Inter-American Observatory, 
NSF’s NOIRLab; the Bok telescope, Steward Observatory, University of Arizona; and the Mayall telescope, 
Kitt Peak National Observatory, NOIRLab. NOIRLab is operated by the Association of Universities for Research in 
Astronomy (AURA) under a cooperative agreement with the National Science Foundation. Pipeline processing and analyses 
of the data were supported by NOIRLab and the Lawrence Berkeley National Laboratory (LBNL). Legacy Surveys also 
uses data products from the Near-Earth Object Wide-field Infrared Survey Explorer (NEOWISE), a project of the Jet 
Propulsion Laboratory/California Institute of Technology, funded by the National Aeronautics and Space Administration. 
Legacy Surveys was supported by: the Director, Office of Science, Office of High Energy Physics of the U.S. Department 
of Energy; the National Energy Research Scientific Computing Center, a DOE Office of Science User Facility; 
the U.S. National Science Foundation, Division of Astronomical Sciences; the National Astronomical Observatories 
of China, the Chinese Academy of Sciences and the Chinese National Natural Science Foundation. LBNL is managed by 
the Regents of the University of California under contract to the U.S. Department of Energy. The complete acknowledgments 
can be found at https://www.legacysurvey.org/acknowledgment/.
This research has made use of the NASA/IPAC Extragalactic Database (NED),
which is operated by the Jet Propulsion Laboratory, California Institute of Technology,
under contract with the National Aeronautics and Space Administration.
\end{acknowledgements}

\end{document}